\documentclass[12pt,preprint]{aastex}
\def\gsim{\;\rlap{\lower 2.5pt
\hbox{$\sim$}}\raise 1.5pt\hbox{$>$}\;}
\def\lsim{\;\rlap{\lower 2.5pt
\hbox{$\sim$}}\raise 1.5pt\hbox{$<$}\;}
\begin{document}
\title{The effect of turbulent intermittency 
 on the deflagration to detonation transition in SN Ia explosions}
\author{Liubin Pan, J. Craig Wheeler and John Scalo}
\affil{Astronomy Department, University of Texas at Austin}
\email{panlb@astro.as.utexas.edu}
\begin{abstract}
We examine the effects of turbulent intermittency on the
deflagration to detonation transition (DDT) in Type Ia
supernovae. The Zel'dovich mechanism for DDT requires
the formation of a nearly isothermal region of mixed ash
and fuel that is larger than a critical size. We primarily 
consider the hypothesis by Khokhlov et al. and Niemeyer and Woosley 
that the nearly isothermal,
mixed region is produced when the flame makes the transition 
to the distributed regime. We use two models for the distribution of
the turbulent velocity fluctuations to estimate the probability
as a function of the density in the exploding white dwarf that
a given region of critical size is in the distributed regime
due to strong local turbulent stretching of the flame structure.
We also estimate lower limits on the number of such regions as
a function of density. We find that the distributed regime,
and hence perhaps DDT, occurs in a local region of critical
size at a density at least a factor of $2-3$ larger than
predicted for mean conditions that neglect intermittency. 
This factor brings the transition density to be much larger than the
empirical value from observations in most situations.    
We also consider the intermittency effect on the more stringent conditions for DDT 
by Lisewski et al. and Woosley.  We find that a turbulent velocity
of $10^8$ cm/s in a region of size  $10^6$ cm, required by Lisewski et al., is
rare. We expect that intermittency gives a weaker effect on the Woosley model
with stronger criterion. The predicted transition density from this criterion remains
below $10^7$ g/cm$^3$ after accounting for intermittency using our intermittency models.

\end{abstract}


\section{Introduction}
A successful model for Type Ia Supernova (SNIa) explosions is required 
to produce a deflagration to detonation transition (DDT) by observational 
constraints. A pure deflagration model gives exploding kinetic energy lower 
than observed (Khokhlov 1991; Gamezo et al. 2003; R\"opke \& Hillebrandt 2005) 
and pure detonation leads to overproduction of iron group elements and 
too little intermediate elements (Branch et al. 1982, 1983). The density 
$\rho_{tr}$ at which the transition occurs determines the amount of 
the nickel produced (H\"oflich 1995; 
H\"oflich and Khokhlov 1996; Dominguez, H\"oflich \& Straniero 2001). 
Therefore a prediction of $\rho_{tr}$, consistent with 
the observed nickel production, is essential to a DDT theory for SNe Ia .
 
The mechanism by which the DDT occurs still remains a mystery. 
The most studied candidate is the Zel'dovich mechanism, which  
requires the existence of an almost isothermal region of mixed ash and fuel 
that is larger than a critical size $l_c$ to drive a supersonic shock 
that is sufficiently strong to sweep over the entire star 
(Khokhlov et al. 1997, hereafter KOW; Niemeyer and Woosley 1997, 
hereafter NW). 
One hypothesis is that a nearly isothermal region is produced 
by turbulent preconditioning.   
KOW argued that, to produce an almost isothermal mixture of ash and 
fuel, the laminar flame must be quenched by turbulent 
stretching, at least locally. This might allow the cold 
fuel to mix with the ash both thermally by electron conduction 
and chemically by diffusivity without being burned. 
They assumed that the criterion to quench a flame is that the turbulent 
velocity at the laminar flame thickness must be larger than the laminar 
flame speed. NW gave a similar argument based on the distributed flame 
burning regime in turbulent combustion. The criterion for a distributed 
flame is expressed in terms of the Gibson scale at which the turbulence 
velocity equals the laminar flame speed. If the Gibson scale is 
smaller than the laminar flame thickness, 
turbulent stretching can generate structures
within the flame and the flame is in the distributed regime.   
NW speculated that in this regime flames can be 
temporally quenched in some regions, which can host the detonation 
after being homogenized in temperature and composition by turbulent 
mixing. The criterion for the distributed regime is equivalent to that for 
flame quenching used by KOW. Both criteria give the same condition on 
the turbulence intensity for given laminar flame properties (see \S 2). 
As the density in the star drops due to the overall expansion, it is easier 
for turbulence to affect the laminar flame because of the decrease in the 
flame speed and the increase in flame thickness. With presumed turbulence 
parameters, the criterion for the turbulence intensity, determined by the  
robustness of laminar flames disturbed by turbulent motions,   
translates to a transition density $\rho_{tr}$ for the DDT. 
         
Several uncertainties exist in the simple model given by these two early 
studies. First, it is not clear whether the criterion used by KOW, 
equivalent to that for a distributed regime (NW), is sufficient for 
flame breaking. How, or even if, flames are quenched  is 
still an open question. Second, it is uncertain whether (local) flame 
quenching is indeed necessary to produce a nearly isothermal region.   
Finally, later studies by Lisewski, Hillebrandt and Woosley (2000) (hereafter
Lisewski et al 2000; see also Lisewski et al. 2000b) and Woosley (2007) find 
that entering the distributed regime, while probably a necessary condition, 
is not sufficient for the DDT to occur. Based on a requirement for 
turbulent transport to be efficient at producing a shallow temperature and 
composition gradient around the laminar flame, Lisewski et al. (2000) find 
that the turbulent velocity at the scale 10$^6$ cm needed for a detonation 
is very large, $\sim 10^8$ cm/s. Woosley (2007) claims that the DDT occurs 
only when the turbulent flame thickness exceeds a critical length scale. 
We show in 
\S 2 that the two criteria, although arising from different physical 
considerations, are basically equivalent. 
The corresponding condition is more stringent than that assumed by KOW and NW.  

In this paper we examine the effect of turbulent intermittency on the 
onset of distributed burning that may relate to the DDT.  
Despite the uncertainties listed above, we will mainly consider the 
model by KOW and NW and use it to illustrate the potential importance of 
intermittency in SN Ia explosions. Our calculations can be applied to the 
criteria by Lisewski et al. (2000) and Woosley (2007) in a straightforward 
way. A quantitative analysis 
using their criteria requires data for laminar flame 
properties and critical length scales at densities below 
$10^7$ g cm$^{-3}$ that are not immediately available (see \S2). 
We give a qualitative discussion of the intermittency effect on their DDT 
models.     

Intermittency is an important concept in turbulence theory. It is characterized 
by intense local events, e.g., strong stretching at small scales, which 
occur at a frequency much larger than predicted from a Gaussian distribution 
(see, e.g., Frisch 1995). The physical origin of  
intermittency in turbulent flows 
is the spatial inhomogeneity in the energy dissipation rate: 
most kinetic energy is viscously dissipated in the finest structures, 
e.g., vortex tubes, which occupy only a small volume fraction. These rare but 
intense dissipative structures give rise to a spatially inhomogeneous and 
intermittent distribution for the turbulent intensity and the stretching rate. 
Intermittency is shown as broad exponential tails in the probability 
distribution for the stretching rate or the dissipation rate at small 
scales (see \S 3). 
The tails get broader at smaller scales, meaning that the probability of 
finding an extreme turbulent stretching rate or intensity increases with 
decreasing scales.   
     
According to 1D simulation results by KOW and NW, the critical size, $l_c$, 
of the isothermal region required for a DDT via the Zel'dovich
mechanism is much smaller, especially at large densities, 
than the expected integral length scale for the buoyancy-driven turbulence 
in SN Ia explosions. This suggests that only a small flame region 
with a sufficiently strong local turbulence intensity may be 
needed to trigger a detonation. Turbulent intermittency, which indicates the 
existence of regions of small sizes where the turbulent stretching is 
much larger than the average value over the flow, is therefore 
expected to have important consequences for DDT. The transition could 
happen earlier at a higher transition density $\rho_{tr}$ than predicted 
by models using the average turbulent intensity. 
At higher densities, much larger turbulent intensity
is required for the DDT, but the rapid decrease of $l_c$ with increasing 
density makes an earlier DDT possible for two reasons. First, the probability 
is larger to find regions of smaller sizes $l_c$ with extreme turbulent 
stretching rate or intensity. Second, there are more regions of smaller 
size available as candidates to host the detonation. Clearly, the 
intermittency effect accounts for the intuitive dependence of 
$\rho_{tr}$ on $l_c$: the smaller the critical size, the easier it may be 
for the transition to happen. To what degree the intermittency effect increases $\rho_{tr}$ is 
the main question we investigate in this paper. 
   

In \S 2, we review the criteria for the DDT in models by KOW, NW, Lisewski 
et al. (2007) and Woosley (2007) and formulate a new criterion taking into 
account the effect of intermittency. We describe two intermittency 
models by Oboukhov (1962) and Kolmogorov (1962) and by She and Leveque 
(1994) in \S 3. Using the intermittency models, we evaluate the transition 
density from the new criteria in \S 4. Our results are summarized and 
discussed in \S 5.

\section{Criteria for the DDT}
The criterion used in NW for judging whether a flame is in the distributed 
regime, which was also assumed to be the condition for the DDT, is to 
compare the Gibson scale $l_G$ with the laminar flame thickness  $l_f$. 
The Gibson scale is defined such that $\delta u(l_G)=S_l$ where 
$\delta u(l)$ is the amplitude of the velocity fluctuations   
at the scale $l$ (or equivalently the velocity difference over a scale 
$l$, i.e., $\delta u(l)=u(l+x)-u(x)$) and $S_l$ is the laminar flame speed
\footnotemark\footnotetext{Note that NW, accounting for the cellular stabilization effect against 
instabilities, e.g., the Landau- Darrieus instability, defined $l_G$ as the 
scale where the turbulent velocity exceeds the effective cellular flame speed. 
This does not introduce a significant difference in the estimate of $l_G$ 
since the effective cellular speed is close to the laminar speed and has 
a very weak dependence on scale; see their Fig. 1.}. If $l_G \gsim l_f$, the 
turbulence cannot internally disturb the flame and the turbulence in effect 
wrinkles the flame. This is called the flamelet regime.    
Only when $l_G \lsim l_f$, 
can turbulence stretch the flame efficiently to generate structures within 
the flame and the turbulent combustion enters the 
distributed regime. 
The condition $l_G \lsim l_f$ is equivalent to $\delta u(l_f) \gsim 
\delta u(l_G)= S_l$ since $\delta u(l)$ is an increasing function of 
the scale $l$\footnotemark\footnotetext{This is also equivalent to 
the diffusivity criterion by Niemeyer and Kerstein (1997) for the onset of 
distributed regime and flame extinction at Prandtl number larger than 
unity, which is the case for white dwarfs. Their criterion was motivated by the 
observation that, at Prandtl number different from unity, two previous  
criteria proposed for the flamelet breakdown and for the flame quenching, 
using the ratio of the flow viscous length scale to the flame thickness 
and the ratio of the viscous timescale to the reaction timescale, 
respectively, are not equivalent.}. The latter, which means that    
the turbulent velocity fluctuation $\delta u(l_f)$ at the scale 
of the flame thickness $l_f$ is larger than the laminar flame speed, 
is   
the criterion used in KOW for flame quenching and the DDT.   

Following KOW, we introduce a factor of $K \sim 1$ in the criterion 
to account for the uncertainty in the flame breaking mechanism, i.e., 
$\delta u(l_f) \geq K S_l$.  We will consider two values for 
$K$, i.e., $K=1$ and $K=8$ (KOW; For $K=8$, to quench a flame, 
the Gibson scale has to be $K^3=512$ times smaller than the flame width). 
This criterion can also be written in terms of timescales. Noting that the 
turbulent stretching timescale, 
$\tau_t$, at the flame thickness is $\tau_t (l_f)=l_f/\delta u(l_f)$ 
and that the nuclear reaction timescale, $\tau_n$, is related to the flame 
speed 
$\tau_n=l_f/S_l$, the criterion is equivalent to $\tau_t(l_f)<\tau_n/K$, 
i.e., to break the flame the stretching timescale at the flame thickness must
be smaller than the nuclear burning timescale (see Niemeyer and 
Kerstein 1997).          

To apply this criterion, the Kolmogorov (1941) 
scaling $\delta u(l)=\bar {\epsilon}^{1/3}l^{1/3} $ is usually used to 
calculate $\delta u(l_f)$ from the turbulent velocity fluctuations at 
large scales where $\bar \epsilon$ is the average dissipation rate in the 
flow. 
From this scaling, the criterion can be written as (KOW, NW), 
\begin{equation}
\bar {\epsilon}^{1/3}l_f^{1/3}>K S_l
\end{equation}
or 
\begin{equation}
\bar {\epsilon} >K^3 S_l^3/l_f = K^3 \epsilon_f 
\end{equation}
where $\epsilon_f$ is defined as $S_l^3/l_f$. Although 
we use the convenient criterion (2) in terms of the dissipation rate 
in our calculations, the turbulent stretching is more fundamental and 
we will use the concept of the flame stretching in our discussions.

The laminar flame speed and thickness depend on the 
chemical composition and the density (Timmes and Woosley 1992, KOW). In Table 
1, we list the flame speed, the 
flame thickness as a function of density for a white dwarf with half carbon 
and half oxygen, mainly taken from Timmes and Woosley (1992). 
The laminar speed decreases and the thickness increases quickly with decreasing density $\rho$, 
therefore $\epsilon_f$ decreases rapidly with 
decreasing $\rho$ as shown in Table 1. 
The average dissipation rate is estimated to be $\bar \epsilon=U^3/L$ 
where $U$ and $L$ are the characteristic velocity and length scales of the 
turbulence, normally set by motions on the large, driving scale.  
At large scales, the turbulence is driven by the Rayleigh-Taylor 
instability.   
The length scale $L$ might be expected to be about the size, $R_f$, 
of the flame region, $L \simeq R_f \sim 10^8$ cm and the velocity scale to be  
about the Rayleigh-Taylor velocity at this scale  
$U \sim 0.5 \sqrt{g_{eff} L}\simeq 10^8$ cm/s where 
the effective gravity is taken to be $ \sim g_{eff}=5 \times 10^8 $ cm/s$^2$
(KOW, NW). Khokhlov (1995), however, showed that 
motions at scales larger than $10^6$ cm freeze out due to the overall 
expansion of the star. In that case, $L\sim 10^6-10^7$ cm and 
$U \sim 10^7$ cm/s. 
We will take $U$ and $L$ as parameters. Note that the criterion eq (2) depends 
on $U$ and $L$ through the dissipation rate. Given the dissipation rate 
$\bar\epsilon$, the critical density below which the inequality (eq 2) is 
satisfied can be obtained using $\epsilon_f$ as a function of $\rho$ 
in Table 1.  For example, if $U \sim 100$ km/s and 
$L \sim 100$ km, $\bar \epsilon \sim 10^{14}$ cm$^2$/s$^3$ and we find that, 
from interpolation in Table 1, 
$\bar \epsilon$ is larger than $\epsilon_f$ at a 
density less than $\sim 4 \times 10^7$ g/cm$^3$. Therefore criterion (2) 
predicts a transition density $\rho_{tr} \simeq 4 \times 10^7$ 
g/cm$^3$ for $K=1$ (see KOW and NW). If $K=8$, the predicted transition 
density is smaller, $\rho_{tr} \sim 1.5 \times 10^7$ g/cm$^3$.  
In the second line of Table 2, we give the predicted $\rho_{tr}$ for
different values of the parameters, which decreases with decreasing 
dissipation rate $\bar \epsilon$.  
The numbers in parenthesis correspond to $K=8$.  

When using the criterion eq (2), we need to keep in mind that the spatial 
fluctuations of $\epsilon$ 
(see \S 3) are completely neglected and the criterion only applies to 
the overall situation in the combustion flow. We will refer to this criterion 
as the mean criterion. When the mean criterion is met, the only implication 
is that the combustion is in the distributed regime in general. 
Considering the intermittency of turbulence, i.e., the spatially 
inhomogeneous distribution of the stretching strength, there can be places 
where the stretching rate is much weaker than the average. These places 
could  still be in the flamelet stage while most other places are in the 
distributed regime. Or conversely, even if the mean criterion (2) is not 
satisfied, one cannot exclude the possibility of there existing a region 
that experiences strong stretching and gets into the distributed regime 
when most of the structure is still in the flamelet regime.   
This latter fact is important for the deflagration to detonation transition. 
The fact that the DDT does not require the entire star to be in the 
distributed regime but instead only needs a region of 
size much smaller than the white dwarf radius (see below, KOW), coupled with 
the intrinsic intermittency, suggests that DDT could occur earlier than 
predicted by eq (2) and hence at a larger transition density. The detonation 
can be triggered locally when a region appears that is larger than the 
critical size and enters the distributed regime due to a strong {\it local} 
stretching.  
It is important to study the degree to which this intermittency effect 
increases the transition density, which is constrained by observations.  
Clearly the answer depends on the critical size, which we consider next.   

The question of how large the isothermal region with well-mixed ash and fuel 
has to be for a detonation was studied by KOW (see also NW). 
In their model, the DDT occurs via the Zel'dovich mechanism (Zel'dovich et al. 
1970) where the mixed region begins spontaneous ignition at the place with the 
minimum induction time, and 
the flame propagates with a phase speed equal to the inverse of the spatial 
gradient of the induction time,  which is large for nearly isothermal and 
well-mixed regions and is not limited by the speed of sound. As the phase 
speed decreases below the Chapman-Jouget speed, a shock forms just ahead 
of the flame front. Whether this shock can explode the whole star depends on 
the strength of the shock when entering the pure fuel, which is determined 
by the size of the isothermal region. If the isothermal region is small and 
the shock is weak, the flame front 
and shock separate with the flame front lagging behind the shock and the 
shock cannot make the whole star explode. The critical strength of the shock 
corresponds to a critical size of the isothermal region, over which the 
shock can be strengthened.  Using 1D simulations, KOW and NW obtained the 
critical size, $l_c$, which depends on the density and the chemical 
composition. It is interesting to note that at early time when the density is 
large, the required size is much smaller than that at later times. We will 
show this has important consequences. The critical size is much smaller than 
what current numerical simulations can resolve, therefore the problem of 
the intermittent stretching at scale $l_c$ cannot be addressed by 
simulations. 

We need a local criterion to check whether a region of a given size $l$, 
in particular $l_c$, is in the distributed regime or not. For that purpose, 
we use a local average dissipation rate $\epsilon_l$ (see eq. 8 in \S 3 for 
a definition) in a region of size $l$ to replace $\bar \epsilon$.  
Following the same argument that leads to equation (2), the criterion for a 
region of size $l$ being in the distributed regime is 
\begin{equation}
\epsilon_l>K^3 \epsilon_f
\end{equation}
where we have used the refined similarity hypothesis by Kolmogorov (1962) 
(see eq. 9 in \S 3). 
Due to the random nature of turbulent flows, $\epsilon_l$ is stochastic, and a 
statistical approach is necessary.  We therefore ask the question: what is 
the probability that any region of size $l$ is in the distributed regime? 
This is given by the cumulative probability 
$P(\epsilon_{l}> K^3 \epsilon_f)$. To answer this question, we need the 
probability distribution  $P (\epsilon_l)$ of $\epsilon_l$.  Fortunately, 
this distribution has been extensively studied in the intermittency models 
for turbulence, which we describe in \S 3. Although these models were 
originally proposed for homogeneous and isotropic turbulence, we will assume 
they apply to SNe Ia where the turbulence is stratified and may not arrive at 
isotropy even at very small scales.    
Once the distribution is specified, one can calculate the probability of 
finding that a region of given size $l_c$ is in the distributed regime, 
\begin{equation}
P(\epsilon_{l_c} >K^3\epsilon_f )=\int \limits_{K^3\epsilon_f}^{\infty}P(\epsilon_{l_c}) d\epsilon_{l_c} 
\end{equation}
which depends on the density through $l_c$ and $\epsilon_f$.   
An immediate examination of eq (4)  shows that, at larger density, the lower 
limit of the integral $K^3 \epsilon_f$ is larger because of the fast flame 
speed and the small flame thickness. This tends to decrease the 
probability.  
However, at larger density, $l_c$ is smaller and the intermittency of 
turbulence tells us that the tail of the distribution $P(\epsilon_l)$ is 
broader for smaller $l$. This tends to counteract the decrease of the 
cumulative probability due to the larger lower integral limit at higher 
densities.   

Furthermore,  for smaller $l_c$, there are more regions of size $l_c$ 
available in the star. This could make the transition occur significantly 
earlier with a transition density considerably larger than predicted by eq (2).
We need to multiply the probability that a given region of size $l_c$ is in 
the distributed regime by the number, $N_{l_c}$, of regions of size $l_c$ 
available in order to calculate the number of regions that are both larger 
than $l_c$ and in the distributed regime at any given density. We assume 
that the deflagration to detonation transition happens 
when 
\begin{equation} 
N_{l_c} \times P(\epsilon_{l_c}> K^3 \epsilon_f )=1.
\end{equation}
Since we are concerned with the flame being stretched into the distributed 
regime, only locations around the flame front are 
of interest when calculating $N_{l_c} $. Therefore, we only count regions 
in the vicinity of the flame front. $N_{l_c}$ depends on the size, $R_f$, of the flame 
region and the flame geometry.
A typical value for $R_f$ is $10^8$ cm (Khokhlov 1995), which could be smaller 
at an earlier time. We will set $R_f \simeq L$ in our calculations in order 
to decrease the number of parameters. Note that 
$R_f >L$ when the freezeout effect is considered and therefore the 
number $N_{l_c}$ we use is a lower limit.     
If the flame region is a 2D spherical front, $N_{l_c} \sim 
4 \pi R_f^2/l_c^2$.  
If the flame structure is highly convoluted, it may have a 
fractal dimension larger than 2. In that case, $N_{l_c}$ is larger. 
The upper limit for $N_{l_c}$ is $\simeq  4 \pi R_f^3/ 3 l_c^3$, which 
applies if the flame geometry is close to 3D.    
Again, we take the lower limit $N_{l_c}=4 \pi R_f^2/l_c^2$, thus 
the transition density we will get is a lower limit.   


As discussed in the Introduction, Lisewski et al. (2000) and Woosley (2007) 
find that entering the distributed regime is not sufficient for the DDT to 
occur and give criteria stronger than that used in KOW and NW.
Lisewski et al. (2000) considered how turbulent transport affects the 
temperature and composition profile around a laminar flame. 
They assumed that, at any point, turbulence translates the temperature 
and composition by a distance $l_t$, over which turbulence can 
transport during a local induction time $\tau_i$.  
The distance $l_t$ is a function of position since $\tau_i$ depends on local 
temperature and composition. It is estimated by the length scale of a 
turbulent eddy with turnover time equal to $\tau_i$, i.e., 
$l_t/\delta u(l_t)=\tau_i$. Using the Kolmogorov (1941) scaling, we get
$l_t=\bar \epsilon^{1/2} \tau_i^{3/2}$.   
For given turbulence intensity, temperature and composition profiles around a 
laminar flame front can be calculated from the translation. 
Clearly, more efficient turbulent 
transport gives shallower temperature and composition profile, 
which is needed for detonation.      
By checking whether the resulting profiles, as initial conditions to 
solve the 1D hydrodynamic equations, can lead to a detonation, 
Lisewski et al. (2000), obtained a condition for the DDT 
on the turbulent intensity. They found 
that, for a successful detonation, the turbulent velocity has to be 
$\sim 10^8$ cm/s at the scale $10^6$ cm. This condition is stronger than 
just entering the distributed regime\footnotemark\footnotetext
{This condition can be converted into a form that can be directly compared 
with eq. 2. Roughly speaking, the physical condition for a detonation in 
this model is that $l_t$ at the laminar flame front is larger than $l_c$, 
i.e., a shallow temperature gradient can be produced over a critical size 
around the flame front. Requiring $l_t>l_c$, we get the criterion 
$\bar \epsilon > (l_c/l_f)^2 (\tau_n/\tau_i)^3 \epsilon_f$. 
Considering that $\tau_i$ defined by Lisewski et al. (2000) is smaller 
than the nuclear timescale $\tau_n$ and that $l_c \gg l_f$, this condition 
is much stronger than the condition of eq. 2. Note that this condition is 
similar to the criterion of Woosley (2007) given below.}. 
Since the expected turbulent velocity at scale $10^6$ cm is $10^6-10^7$ cm/s, 
Lisewski et al. (2000) concluded that a DDT via the Zel'dovich mechanism in 
SNe Ia is unlikely. However, considering the spatial inhomogeneity of 
turbulent intensity, i.e., intermittency, it is possible for regions of size 
$10^6$ cm with large enough turbulent velocity to arise.    

The result of Lisewski et al. (2000) motivated R\"opke (2007) to study the 
probability of finding a region of size $10^6$ cm with a turbulent rms 
velocity of $\sim 10^8$ cm/s. Using data from 3D numerical simulations with a 
turbulent subgrid-scale method, R\"opke (2007) analyzed the velocity 
fluctuations at the grid size ($10^6$ cm) and obtained a fat exponential 
tail for large velocity fluctuations that extends up to $10^8$ cm/s. The large 
velocity fluctuations seem likely to be located at the trailing edge of a 
bubble-like feature (R\"opke 2007). 
This confirms the intermittency in the turbulent combustion flow in SNe Ia; 
there exist grid cells where the turbulent intensity is much stronger 
than the average. From the probability of finding a grid cell with required  
turbulent intensity, R\"opke concluded that the DDT 
triggered by a local cell with large velocity fluctuations is possible 
but probably rare. In our notations, the probability is given by 
$P(\epsilon_{10^6 \hspace{0.5mm} {\rm cm}}> 10^{18}\hspace{1.5mm} 
{\rm cm^2/s^3})$ where $10^{18}$ cm$^2$/s$^3$ corresponds to the
dissipation rate in a region of size $10^6$ cm with a rms velocity 
of $10^8$ cm/s. We will calculate this probability and consider the availability
of such regions using two intermittency models given in \S 3 and 
compare with the results of R\"opke (2007) in \S 4.    
  
Woosley (2007) proposed a new criterion for the DDT based on a 
calculation of the distributed flame width using an eddy diffusivity 
approximation. Making an analogy to the estimate of the laminar flame 
thickness, Woosley (2007) obtained the distributed flame width 
$\lambda$ from the equation $\lambda \simeq (D(\lambda) \tau_n)^{1/2}$ 
where $D(\lambda)= \delta u(\lambda)\lambda$ is the eddy diffusivity at 
scale $\lambda$ and $\tau_n$ is the nuclear reaction timescale. 
Using the Kolmogorov (1941) scaling for $\delta u(\lambda)$,  
the distributed flame width is given by 
$\lambda=\bar \epsilon^{1/2} \tau_n^{3/2}$ (note that this formula for 
$\lambda$ is similar to $l_t$ in Lisewski et al. (2000)).  
Woosley (2007) assumed that the condition for detonation is that the 
minimum burning timescale in the distributed flame is smaller than 
the sound crossing time over the distributed flame width $\lambda$, 
or equivalently, $\lambda \gsim r_{sonic}^{min}$ where 
$r_{sonic}^{min}$ is the sound crossing length over the minimum burning 
timescale in the distributed flame. The minimum sound crossing 
length is thus 
the critical size of the distributed flame width for detonation. 
The criterion $\lambda \gsim r_{sonic}^{min}$ is equivalent to 
$\bar \epsilon \gsim (r_{sonic}^{min})^2/\tau_n^3$. 
Noting that $\tau_n=l_f/S_l$, the condition can be written as
$\bar \epsilon \gsim (r_{sonic}^{min}/l_f)^2 \epsilon_f$. 
Since $r_{sonic}^{min}$ given in Table 4 of Woosley (2007) is close to 
$l_c$ listed in Table 1 in this paper, we will use $l_c$ instead of 
$r_{sonic}^{min}$ for simplicity, i.e.,  
\begin{equation}
\bar \epsilon \gsim (l_c/l_f)^2 \epsilon_f. 
\end{equation}
which is much stronger than eq (2) because $l_c$ is much larger than 
the laminar flame thickness $l_f$. This condition can be used to determine 
the transition density $\rho_{tr}$ by the same calculation process 
as in the case of the criterion eq (2). Note that, except a factor 
of $(\tau_n/\tau_i)^3$, this criterion is basically 
equivalent to that given in footnote (3) for the requirement by Lisewski 
et al. (2000). We find that, for the reasonable  
turbulence parameters listed in Table 2, the criterion results in a 
transition density below $10^7$ g cm$^{-3}$ and we cannot specify it 
due to the lack of data for $l_c$, $S_l$ and  $l_f$ at densities below 
$10^7$ g cm$^{-3}$. 

In his estimate for $\rho_{tr}$, Woosley (2007) used $U=10^8$ cm/s at scale 
$L=10^6$ cm throughout the calculations, based on 
the result by R\"opke (2007) on the possibility of the existence of
regions of size $10^6$ cm with a rms velocity of $10^8$ cm/s.  
With these turbulence parameters he derived $\rho_{tr}=10^7$ g cm$^{-3}$. 
Clearly, in Woosley's calculation, the intermittency effect implicitly 
contributes to the transition density obtained because, as 
discussed earlier, a turbulent rms velocity of $10^8$ cm/s at $10^6$ cm 
can only arise from intermittency.  
 
The intermittency effect for the criterion of Woosley (2007) can be 
included more consistently in our formulation. Instead of considering a 
single special scale $10^6$ cm, our model specifies intermittency over a 
continuous range of scales corresponding to critical sizes at 
different densities. Following the same steps that lead to eq (5), we 
incorporate the intermittency effect in the DDT model of Woosley (2007) 
and obtain a criterion,   
\begin{equation}
N_{l_c}\times  P(\epsilon_{l_c}> (l_c/l_f)^2 \epsilon_f)=1. 
\end{equation}
which only differs from eq. 5 by the lower limit in the cumulative probability.
We will discuss about this criterion in \S 4.  

We point out that the eddy diffusivity method used by Woosley (2007) to 
approximate the combined action of the turbulent advection and the 
microscopic diffusivity is an oversimplification. This procedure 
implicitly assumes a smooth structure in the distributed flame and 
neglects the fluctuations of temperature and concentration, which may 
be important in determining the effective width of distributed flames.
 
\section{Intermittency} 
Kolmogorov's 1941 theory assumes that the energy transfer in the inertial 
range is equal to the average dissipation rate $\bar \epsilon$ in the flow 
and is the same throughout the inertial scales down to the viscous scale 
where the kinetic energy is removed. This assumption, together with the 
similarity hypothesis, predicts that the statistics of the velocity 
difference (or the velocity fluctuations) at any inertial scale 
is completely determined by the average dissipation 
rate $\bar \epsilon$.
However, fluctuations in the dissipation rate clearly exist as can be seen 
from the formula for the local viscous dissipation rate, $\epsilon ({\bf x},t)
=\frac{\nu}{2}\sum\limits_{i,j}(\partial_i u_j+\partial_j u_i)^2$, which is 
a function of the fluctuating velocity field. 
The spatial fluctuations in $\epsilon$ are well-illustrated by the 
intense dissipation 
structures at small scales such as vortex tubes.
This effect needs to be taken into account for a more accurate 
prediction of the scaling behavior of the velocity difference (Landau and Lifshitz 1944). The statistics of the velocity difference over a separation $l$
depends on the distribution of the dissipation rate over regions of 
size $l$, which is defined as (e.g., Kolmogorov 1962),  
\begin{equation}
\epsilon_l ({\bf x},t)= \frac{3}{4 \pi l^3}\int\limits_{|{\bf x'}|<l} \epsilon 
({\bf x}+{\bf x'},t)d{\bf x'}.
\end{equation}
 Clearly, the mean 
of $\epsilon_l$ is equal to $\bar \epsilon$ and thus is independent of $l$. 
This means that the {\it average} energy flux over all the inertial scales 
is constant. The $\epsilon_l$ distribution is essential to the intermittency models for turbulence. Note that this distribution is exactly
what we need in our calculations for the transition of the turbulent 
combustion 
to the distributed regime by turbulent stretching and quenching described in 
\S 2, eqs (4) and (5).   


Intermittency in turbulence is usually expressed in terms of
the scaling behavior of the structure 
functions $\langle \delta u(l) ^p \rangle \sim l^{\zeta_p}$ where 
$\delta u(l)=u(x+l)-u(x)$ is the (longitudinal) velocity difference and 
$\zeta_p$ is the scaling exponent for the $p$th- order structure function.  
Kolmogorov's 1941 theory predicts that the exponent $\zeta_p$ goes with 
$p$ as $\zeta_p =p/3$.   
However experimental data (e.g., Anselmet et al. 1984) have shown departure 
from this linear relation and $\zeta_p$ increases
significantly slower than $p/3$ at large $p$.  
This ``anomalous'' scaling is referred to as intermittency. The data indicate 
broader and broader tails for the distribution of $\delta u(l)$ at 
smaller and smaller scales, e.g., the kurtosis of the distribution, 
$\langle \delta u(l)^4 \rangle/\langle \delta u(l) ^2\rangle^2 \propto l^{\zeta_4-2\zeta_2}$, increases 
with decreasing $l$ because $\zeta_4<2\zeta_2$. The distribution 
of $\delta u(l)$ is fatter for smaller $l$.  
The anomalous scaling is fundamentally caused by the fluctuations in    
the dissipation rate $\epsilon_l$. 
%
Applying the refined similarity argument for homogeneous and isotropic 
turbulence (Kolmogorov 1962), the velocity 
difference over a separation $l$ can be related to the dissipation 
rate $\epsilon_l$,
\begin{equation}
\delta u(l) \sim \epsilon_l^{1/3} l^{1/3}.  
\end{equation}   
(note that the Kolmogorov's 1941 theory uses $\bar \epsilon$.)
The structure functions are then given by, 
\begin{equation}
\langle \delta u(l)^p \rangle \propto \langle \epsilon_l^{p/3} \rangle  l^{p/3}.  
\end{equation} 
Clearly, the departure from the linear scaling for the velocity 
difference comes from the statistics of the dissipation rate.  
Assuming $\langle \epsilon_l^p \rangle \propto l^{\tau_p}$ (e.g., She and Leveque 1994), we have, 
\begin{equation}
\zeta_p=p/3+\tau_{p/3}.  
\end{equation}
Developing a physical model for $\tau_p$ that satisfies the experimental 
result for $\zeta(p)$
has been the main task of intermittency theories. 
Although we are mainly concerned with the distribution of $\epsilon_l$, discussions of the structure functions are necessary because they are directly 
measurable in experiments and give 
important information and constraints on the $\epsilon_l$ distribution.     
We will use two intermittency models in our calculations: the log-normal model 
(Oboukhov 1962 and Kolmogorov 1962) and the log- Poisson model by 
She and Leveque (1994).      

\subsection{The log-normal model}
Oboukhov (1962) and Kolmogorov (1962) developed the first intermittency model. 
In this model, the distribution of $\epsilon_l$ is assumed to be 
log-normal (Kolmogorov 1962). A justification for this ``natural''  distribution for $\epsilon_l$ 
was given by Yaglom (1966). Imagine the cascade progress as 
successive eddy fragmentations from the integral scale $L$ to the dissipation 
scale $\eta$. The statistics of the energy flux at an inertial scale $l$  (or equivalently the dissipation 
rate $\epsilon_l$) depends on the fragmentations before the scale is reached. The total number $N$ of steps that lead to the scale $l$ is proportional to $N \sim ln (L/l)$. Defining  $\chi_i = \epsilon_i/\epsilon_{i-1}$ as the ratio of the energy transfer rates at two successive fragmentation steps,
the  energy flux at the scale $l$ can be expressed in the ratios (see e.g., Monin \& Yaglom 1975), 
\begin{equation}
\epsilon_l= \epsilon_L \chi_1 \chi_2 ...\chi_N       
\end{equation}       
where $\epsilon_L$ is the dissipation rate at the integral scale (or the transfer flux at the largest scale),  $\epsilon_L \simeq \bar \epsilon $.  
Due to  the randomness in the fragmentation process, $\chi_i$'s are stochastic variables.  
Assuming  a self- similar fragmentation process, the distributions of $\chi_i$'s are similar
and $ln (\epsilon_l/\epsilon_L)=\sum\limits_{i=1}^N ln(\chi_i)$ is expected to 
be Gaussian from the central limit theorem,
\begin{equation}
p(\epsilon_l) d\epsilon_l=\frac{1}{\sqrt{2\pi \sigma_l^2}} 
exp(-\frac{(ln (\epsilon_l / \bar \epsilon)+\sigma_l^2/2)^2}
{2\sigma_l^2}) d ln (\epsilon_l /  \bar \epsilon) 
\end{equation}
where the variance $\sigma_l^2$ is proportional to the number of steps 
$\sigma_l^2= \mu ln (L/l)$ with $\mu$ being a parameter to be determined by experimental data 
and the $\sigma_l^2/2$ term in the numerator in the exponential is to guarantee 
the mean $\langle \epsilon_l \rangle$ is equal to the overall average dissipation rate 
$\bar \epsilon$.  
This distribution will be used later to calculate the probability (eq. 4) for a region 
of a given size being in the distributed regime. 

The scaling behavior of $\epsilon_l$ can be derived by integrating equation (13), 
\begin{equation}
\langle \epsilon_l^p \rangle \propto l^{-\frac{1}{2}\mu p(p-1)}
\end{equation}
which gives $\tau_p=-\frac{1}{2}\mu p(p-1)$. From eq (11), we have 
\begin{equation}
\zeta_p=p/3-\frac{1}{18}\mu p(p-3)
\end{equation}
Therefore $\mu=2-\zeta_6$, which can be obtained from the results of 
experiments and simulations.  
It has been found that $\mu \simeq 0.2$ (Frisch 1995, Biskamp 2000). 
The relation (15) agrees with experiments quite well at small $p$ but starts 
to exhibit deviation at 
$p \gsim 10$  and gives an unrealistic maximum and turnover at $p>16$, 
violating the requirement that the $\zeta(p)$- $p$ curve must be monotonic 
and concave (Frisch 1995). Simulations by Wang et al. (1996) 
suggest that this disagreement corresponds to the departure of the 
distribution for $ln (\epsilon_l/\bar \epsilon)$ from normal at scales 
close to the dissipation scale. They find 
that, at these scales, the distribution of $ln(\epsilon_l/\bar \epsilon)$ 
shows a negative skewness, meaning that the log-normal distribution 
overestimates the probability in the very high $\epsilon_l$ tail.
However the distribution of $\epsilon_l$ agrees with log-normal very well in 
the inertial range away from the dissipation scale and the agreement 
is better and better for larger and larger scales (Wang et al. 1996). 
Fortunately the critical scale we are concerned with is well within the 
inertial range (see \S 4) and according to Fig. 6 in Wang et al. (1996) 
the log normal fit is very good at least up to the $4-\sigma$ tail. 
They also show that the fit gets better as the Reynolds number
increases. More recent simulations by Yeung et al. (2006) with resolutions 
up to $2048^3$ obtained similar results. The log normal distribution gives
a very good fit to $4- \sigma$ and only deviates by a factor of $2$ at 
the $5 -\sigma$ tail. They also find that the negative skewness gets closer to 
zero with increasing Reynolds number. The Reynolds number in SNe Ia is 
$Re \sim 10^{14}$ for typical velocity scale $10^7$ cm/s, length 
scale $10^7$ cm and viscosity $1$ cm$^2$/s. This is much  
larger than in all the current simulations. 
Therefore one may expect that the log-normal 
distribution probably applies even further out on the tail for the 
inertial scales of 
the turbulence in SNe Ia.  
However, the departure of the predicted $\zeta(p)-p$ curve from 
the experiments (with high Re) at $p \gsim 10 $ suggests that, even 
at huge Reynolds number, the log normal fit eventually breaks down at some 
large $\epsilon_l$ in the tail. 
Therefore we need to be careful when using the log-normal 
model. We will give more discussion on this point
in the calculations given in \S 4.    

Another issue is that the distribution of $\epsilon_l$ has a physical 
cutoff in a realistic system due to the finite viscosity. 
Since the intermittency is stronger at smaller scales, 
the cutoff in the distribution of $\epsilon_l$ is probably larger for 
smaller $l$ and obtains a maximum at the dissipation scale, $\eta$. 
For Kolmogorov scaling, the cutoff in the distribution of $\epsilon_\eta$ is 
is given by $\bar{\epsilon} Re^{1/2}$. 
Since $Re \sim 10^{14}$ in SNe Ia, this maximum dissipation rate is far beyond 
that required to break flames at density $\lsim 10^8$ g/cm$^3$. 
Therefore ignoring this maximum cutoff does not affect our result.  
However, the largest available dissipation rate at 
an inertial scale $l$ is probably smaller than the cutoff in the distribution
of $\epsilon_\eta$ and thus may affect the calculation for the cumulative probability 
defined in eq (4) if the cutoff in the distribution $P(\epsilon_{l_c})$ is close to 
or even smaller than $K^3\epsilon_f$. Since the log-normal model does not 
address the cutoff in the distribution of $\epsilon_l$, we will neglect this 
potential effect in this model. 

On the other hand, the log-Poisson model  
we consider in the next section gives a maximum dissipation rate at each 
inertial scale, corresponding 
to the strongest dissipative structures at that scale. In that model, 
a nonzero cumulative probability in eq (5) requires the lower limit $K^3\epsilon_f$ in eq (4) be 
smaller than the maximum.

\subsection{The log-Poisson model}

A major success in the intermittency theory is the model by She and 
Leveque (1994).  
In this model, She and Leveque studied the hierarchy of dissipation 
intensity in structures of size $l$ and, by invoking an unknown 
``hidden symmetry,'' they related the characteristic dissipation 
rates in structures of different intensity levels to the strongest 
dissipative structures. This relation gives a prediction of  $\tau_p$ 
as a function of $p$, which only depends on the proprieties of the most 
intermittent structures. Assuming that the dissipation rate in 
regions of size $l$ containing the most intense structures 
exhibits a scaling  $\propto l^{-2/3}$ (see explanation in Appendix A) 
and the most intermittent structures are filamentary, corresponding to a 
codimension of 2, She and Leveque obtained a $\zeta(p)-p$ relation, 
which is in excellent agreement with experimental data. 
The ``hidden symmetry''  has been immediately interpreted as  
a log-Poisson process (Dubrulle 1994, She and Waymire 1995) in a 
multiplicative cascade model. In this section, we adopt the log-Poisson 
version of the She-Leveque model. The original presentation by She and 
Leveque (1994) is given in Appendix A. 

In a multiplicative model, the dissipative rates at two 
scales $l_2$ and $l_1$ ($l_1>l_2$) are related by a 
multiplicative factor $W_{l_1l_2}$, 
\begin{equation}
\epsilon_{l_2}=W_{l_2l_1} \epsilon_{l_1}
\end{equation}  
The average $\langle W_{l_1l_2} \rangle$ is equal to unity 
since $\langle \epsilon_{l_1} \rangle=
\langle \epsilon_{l_2} \rangle=\bar \epsilon$.   
She and Waymire (1995) speculated that $W_{l_2l_1}$ consists of two events. 
First is the amplification of the dissipation rate in the cascade, 
which tends to produce singular structures with 
$\epsilon_{l_2} \propto (l_1/l_2)^\gamma$ approaching infinity as $l_2$ 
goes to $0$. The meaning of $\gamma$ is discussed below. To ensure 
$\langle W_{l_1l_2} \rangle=1$, a second event is required to 
reduce $W_{l_1l_2}$. She and Waymire (1995) called this event the
modulation-defects since it modulates the singular structures. The 
defects were assumed to be a discrete Poisson process. Each of the defects 
decreases $W_{l_1l_2}$ by a factor of 
$\beta$, thus 
\begin{equation}
W_{l_2l_1}=(l_1/l_2)^\gamma \beta^n
\end{equation}
if there are $n$ defect events in the cascade. 
The number $n$ of the events that occur in the cascade from the scale 
$l_1$ to $l_2$ obeys a Poisson distribution, 
\begin{equation}    
P(n)=exp(-\lambda_{l_1l_2}) \frac{\lambda_{l_1l_2}^n}{n!}
\end{equation}
where $\lambda_{l_1l_2}$ is the mean number of the defect events in the 
cascade, which 
is expected to be proportional to the total number of the cascade steps, 
i.e.,  
$\lambda_{l_1l_2} \propto ln(l_1/l_2)$. In fact, $\lambda_{l_1l_2}$ can 
obtained by taking the average of eq. (17)
and requiring $\langle W_{l_1l_2} \rangle=1$. 
Using the identity $\sum\limits_{n=0}^\infty \frac {\alpha^n}{n!}=exp(\alpha)$,
we get $\langle \beta^n \rangle=exp((\beta-1)\lambda_{l_1l_2})$ for the Poisson
distribution eq (18), therefore,   
\begin{equation}
\lambda_{l_1l_2}=\frac {\gamma ln(l_1/l_2)}  {1-\beta}
\end{equation}  

In this model, there is a largest dissipation rate at any scale. 
Clearly the largest 
dissipation rate is achieved if there is no defect, i.e., $n=0$, 
in a cascade from the integral scale $L$ to the scale $l$ of interest, thus 
the largest dissipation rate is equal to $\epsilon_L (L/l)^\gamma $. 
This largest dissipation rate 
corresponds to $\epsilon_l^{(\infty)}$ in Appendix A. 
Similarly  $n=1$ gives the second strongest dissipative rate at a given scale, 
and so on.       

From eqs (16) and (17), we have, 
\begin{equation}
ln(\epsilon_{l_2}/\bar \epsilon)=ln(\epsilon_{l_1}/\bar \epsilon)+\gamma 
ln(l_1/l_2)+nln(\beta)
\end{equation} 
thus, using the Poisson distribution for $n$, the distribution for the 
dissipation rate at $l_2$ can be derived from that at any scale $l_1$ 
larger than $l_2$. 
In particular,  
we consider deriving the distribution of $\epsilon_l$ at any scale $l$ from 
the integral scale
$L$. The distribution function of $\epsilon_L$ at the integral scale 
depends on how the energy is injected in the flow, thus is not universal and 
may vary from flow to flow. Therefore the function form cannot be specified. 
However, there is a strong constraint for its width. 
Since $\epsilon_L \simeq \bar \epsilon$, the distribution of 
$ln(\epsilon_L/\bar \epsilon)$ is expected to be very narrow around 
$ln(\epsilon_L/\bar \epsilon) \simeq 0$ and hence to be approximately 
a delta function. We denote the distribution of $ln(\epsilon_L/\bar \epsilon)$
as $P_L( ln(\epsilon_L/\bar \epsilon))$. It then follows from eqs (18) and
 (20) that 
\begin{equation}
P(\epsilon_l) d\epsilon_l=\sum\limits_{n=0}^{\infty} exp(-\lambda) 
\frac{\lambda^n}{n!}P_{L}(ln(\epsilon_{l}/\bar \epsilon)- \gamma ln(L/l)- n ln(\beta)) dln(\epsilon_l/\bar\epsilon)
\end{equation}
where $\lambda=\lambda_{Ll}=\gamma ln(L/l)/(1-\beta)$.
Each term in eq (21) represents the contribution from dissipation structures 
of different levels, e.g., the $n=0$ term corresponds to the most intensive 
structures of size $l$. 

To compare the model with experiments and obtain the parameters, we calculate 
the moments $\langle \epsilon_l \rangle$ from the distribution eq. (21), 
\begin{equation}
\begin{array}{lll}
{\langle \epsilon_l^p \rangle}
&\propto& \sum \limits_{n=0}^\infty \int exp(-\lambda) \frac {\lambda^n}{n!}
exp(px) P_L(x-\gamma ln(L/l)- n ln(\beta))dx\\
 &=& (l/L)^{-\gamma p} exp(-\lambda) \sum \limits_{n=0}^\infty \frac {(\beta^p \lambda)^n}{n!} 
 \int exp(x') P_L(x')dx' \\
 &=& B_p exp(-\lambda(1-\beta^p)) (l/L)^{-\gamma p}\\
 &=&B_p (l/L)^{-\gamma p+\gamma (1-\beta^p)/(1-\beta)}
 \end{array}
 \end{equation}
where  we used a variable change $x'=x-\gamma ln(L/l) - n ln(\beta)$ in the second step and the identity 
$\sum\limits_{n=0}^\infty \frac {\alpha^n}{n!}=exp(\alpha)$ in the third step.
The coefficients $B_p=\int  exp(px') P_L(x') dx' $; $B_0=B_1=1$ from the normalization of $P_L$ and the requirement that
$\langle \epsilon_L \rangle=\bar \epsilon$ respectively. 

The result eq (22) gives $\tau_p=-\gamma p+\gamma (1-\beta^p)/(1-\beta)$, which
is the same as (A8) in Appendix A, meaning that the ``hidden symmetry" 
described in the appendix is equivalent to a log-Poisson process.   
The parameters $\gamma$ and $\beta$ introduced here are identical to those 
described in the appendix, and thus have the physical meanings explained 
there, i.e., $\gamma$ can be interpreted as the exponent of the dissipation 
rate scaling in regions containing the most intermittent structures and 
$\beta$ is related to the codimension $C$ of the strongest dissipation 
structures, $\gamma/(1-\beta)=C$ (see Appendix A for details). 
As discussed in the appendix, She and Leveque argued that $\gamma=2/3$ and 
$\beta=2/3$ for $C=2$ corresponding to filamentary dissipation structures 
in incompressible turbulence. This results in $\zeta_p$ as a function of 
$p$ that agrees with the experiments with an accuracy of $1\%$, implying eq 
(21) provides a good distribution for $\epsilon_l$. 
The She-Leveque formulation has been extended to supersonic 
turbulence (Boldyrev et al. 2002) and MHD turbulence (Muller \& Biskamp 2000) 
where the dissipation structures are dissipation sheets and the current 
sheets, respectively. For these 2 dimensional dissipation structures, 
the codimension $ C =1$ and $\beta=1/3$.         
In next section we use the log-Poisson distribution (eq 21) in our 
calculations for the cumulative probability in eq 4. We will take $\gamma=2/3$ and consider both  
filaments ($\beta=2/3$) and sheets ($\beta=1/3$) as the most intermittent 
dissipation structures.    
 

\section{Results}
\subsection{The log-normal model}
We are ready to calculate the probability $P (\epsilon_{l_c}>K^3 \epsilon_f)$ 
using the distributions $P(\epsilon_l)$ given in \S 3. The calculation is 
straightforward for the log-normal distribution eq. (13), 
\begin{equation}
\begin{array}{lll}
P(\epsilon_{l_c}>K^3\epsilon_f) &=& \int \limits_{ln(K^3 \epsilon_f/ \bar \epsilon)}^{\infty}
\frac{1}{\sqrt{2 \pi \sigma_{l_c}^2}} exp(-\frac {(x+ \sigma_{l_c}^2/2)^2} {2\sigma_{l_c}^2}) dx \\
&=& \int\limits_{\frac {ln(K^3\epsilon_f/ \bar \epsilon)+\sigma_{l_c}^2/2}{\sigma_{l_c} }}^{\infty}
\frac{1}{\sqrt{2 \pi}} exp(-x^2/2) dx\\
&=&\frac{1}{2} erfc(\frac {ln(K^3\epsilon_f/ \bar {\epsilon}) } 
{\sqrt{2} \sigma_{l_c}}+\frac{\sigma_{l_c}}{2\sqrt{2}} )
\end{array}
\end{equation} 
where $\bar\epsilon \simeq U^3/L$, $\sigma_{l_c}^2=\mu ln(L/l_c) \simeq 0.2 ln(L/l_c)$ and $erfc(x)$ is the complementary error function.
Using $\epsilon_f$ and $l_c$ given in Table 1, we calculated the probability as a function of 
the density assuming different values for the characteristic velocity ($U$) and length ($L$) scales. For example, if $U=100$ km/s and $L=100$ km, the 
probabilities are $0.5 \times 10^{-10}$, $0.042$, $0.9$ and $1$ 
at $\rho=10^8$, $5 \times 10^7$, $3\times 10^7$ and $10^7$ g/cm$^3$
respectively if $K=1$. It is interesting to note that, 
at a density $3 \times 10^7$ g/cm$^3$, 10\% of the local regions of 
the critical size are still in the flamelet regime, although generally the 
flame has reached the distributed regime according to the mean criterion 
eq (2). With $R_f \simeq L = 100$ km, at the 4 densities above 
from high to low, the corresponding numbers $N_{lc}= 4 \pi R_f^2/l_c^2$ of 
regions of the critical size that cover the flame front 
are $3\times 10^{10}$, $7 \times 10^8$, $5 \times 10^7$ and $3 \times 10^4$. 
Multiplying $P(\epsilon_{l_c}>K^3\epsilon_f)$ with $N_{l_c}$ (eq. 5), we see that 
there is already one region of size $l_c$ in the distributed regime at a 
density $10^8$ g/cm$^3$. Recalling that, according to the criterion eq (2), 
the DDT does not occur until the density decreases to $4 
\times10^7$ g/cm$^3$, we find that in this case 
the intermittency effect may increase the transition density by more than 
a factor of 2. 
   
We point out the cumulative probability calculated from eq (23) at 
density $10^8 $ g/cm$^3$ in the example above comes from a little beyond 
the $6-\sigma$ tail of the distribution for $ln(\epsilon_l/\bar\epsilon)$ 
and we need to check whether the log-normal distribution there is a good 
approximation. As discussed in \S 3.1, numerical simulations have shown that 
for a scale $l$ in the inertial range, the distribution of $\epsilon_l$ is well 
approximated by log-normal up to the $5-\sigma$ tail (Yeung et al. 2006). 
Assuming the Kolmogorov scaling, the dissipation scale in Type Ia SNe is    
$\eta =L Re^{-3/4} \simeq 10^{-3}$ cm for $L\sim 10^7$ cm and the 
Reynolds number $Re \simeq 10^{14}$. The critical scale 
$l_c \sim 10^2-10^4$ cm of interest here is well between the integral scale 
and the dissipation scale, thus we expect that the 
distribution for $\epsilon_{l_c}$ is close to log-normal at least up to 
the $\sim 5-\sigma$ tail. The question is then 
whether the good fit extends further.     
Wang et al. (1996) and Yeung et al. (2006) found that the 
log-normal fit is better for larger Reynolds number, thus 
it is expected that the log-normal approximation probably applies to higher 
on the tail than $5-\sigma$. 
As argued in \S 3.1, the log normal approximation eventually fails 
somewhere in the extreme tail even at high Reynolds number.  
To know exactly how far the log-normal fit extends, numerical 
simulations with much higher resolution are needed.
We have to be careful about the validity of the log-normal approximation in 
the far tail because it overestimates the probability distribution 
for $\epsilon_l$ once it breaks down and in that case eq (23) 
overestimates $P(\epsilon_{l_c}>K^3\epsilon_f)$.  

Due to the complication of the validity of the log-normal distribution at 
the far tail, we consider two extreme cases and give the upper and lower 
limits for the transition density.      
First, we ignore the departure from log-normal and  
evaluate the density at which $N_{l_c} \times P(\epsilon_{l_c}>K^3\epsilon_f)=1$ 
using eq (23) for $P(\epsilon_{l_c}>K^3\epsilon_f)$ and Table 1 for $\epsilon_f$ and $l_c$ with 
different parameters $U$ and $L$. We will denote this density as $\rho_{LN}$ 
with the subscript $LN$ standing for log-normal. Interpolation was used to 
obtain $\epsilon_f$ and $l_c$ not tabulated in Table 1. If the distribution 
of $\epsilon_l$ is exactly log-normal as given by eq (13), then $\rho_{LN}$ 
is the predicted transition density for the DDT with the intermittency taken 
into account. On the other hand, if the log-normal distribution 
overestimates the probability at the high tail, eq (23) overestimates 
the cumulative probability and $\rho_{LN}$ is the upper limit for $\rho_{tr}$.  
We give $\rho_{LN}$ for different parameters $U$ and $L$ in the second line of 
Table 2. 


In the other extreme, we assume that the log-normal distribution fails to 
fit the distribution of $\epsilon_l$ beyond the $5-\sigma$ tail. This gives a 
lower limit for the transition density since numerical simulations have shown 
that the log-normal fit applies at least to $5-\sigma$. In this case, 
we keep track of the integral limit in the second line of eq (23) 
at $\rho_{LN}$, which tells us which part 
of the tail of the distribution gives the main contribution to $P(\epsilon_{l_c}>K^3\epsilon_f)$ 
at that density. If the integral limit is smaller than $5$, the contribution 
to the cumulative probability  
is from within $5-\sigma$ and vice versa.  
We calculate the density at which the integral limit is equal to $5$ and
denote this density as $\rho_{5\sigma}$. Since the integral limit is a 
decreasing function of the density, if $\rho_{LN}<\rho_{5\sigma}$, 
the contribution to $P(\epsilon_{l_c}>K^3\epsilon_f)$ at density $\rho_{LN}$ is from within $5-\sigma$. 
In this case, the cumulative probability calculated from eq (23) is valid 
and $\rho_{LN}$ is a good estimate for the transition density. Otherwise 
if $\rho_{LN}>\rho_{5\sigma}$, the contribution to the probability 
is from beyond the $5-\sigma$ tail, eq (23) overestimates it and thus $\rho_{LN}$ 
overestimates the transition density $\rho_{tr}$. In this case, 
$\rho_{5\sigma}$ gives a lower limit for the transition density because  
at $\rho_{5\sigma}$, we have $N_{l_c} \times P(\epsilon_{l_c}>K^3\epsilon_f)\gg 1$ 
using eq. (23) which applies for $\rho \leq \rho_{5\sigma}$.  
Therefore if the log-normal fit fails just beyond $5-\sigma$, we have 
a lower limit for the transition density, $min(\rho_{LN},\rho_{5\sigma})$.   
We give this lower limit in the 3rd line of Table 2.  


Similar calculations can be done for the $K=8$ case. The results of the 
upper and lower limits for the transition density in the $K=8$ case are given 
in parenthesis in Table 2. Comparing with predictions from the mean criterion 
eq (2) (the first line in Table 2), the log-normal model predicts that the 
intermittency effect increases the transition density by a factor of 2-3 
for all the cases we list in Table 2.

We evaluate the probability of the existence of a region of size $10^6$ cm with 
a rms velocity $10^8$ cm/s, required for the DDT by Lisewski et al. (2000), 
and compare with the numerical results of R\"opke (2007). Using the log-normal 
distribution for $\epsilon_l$, we find that the requirement 
$P(\epsilon_{\rm 10^6 \hspace {1mm} cm}>10^{18} 
\hspace{1.5mm} {\rm cm^2/s^3})$
requires conditions from the extreme tail of the distribution. The
likelihood is completely negligible ($\sim 10^{-40}$) if the velocity 
$U$ at the integral length scale is less than $\sim 10^7$ cm/s. 
Only if $U$ is larger than $5 \times 10^7$ cm/s is the probability 
appreciably larger so that the required region might be available. 
For example, if $U=5 \times 10^7$ cm/s at $L=10^7$ cm, the probability 
is $\sim 10^{-11}$. This is still too small to guarantee the 
existence of a region as required by Lisewski et al. (2000). 
The number of available candidate regions of size $10^6$ cm around the 
flame front is probably smaller than $10^5-10^7$, assuming the 
flame front radius is $\sim 10^8-10^9$ cm.  This result agrees 
with the conclusion of R\"opke (2007) that the existence of a region 
as required by Lisewski et al. (2000) is rare. To ensure such a region, 
the velocity at the integral scale has to be larger than $10^8$ cm/s, 
which is probably impossible as discussed in \S 2. 

We also carry out a calculation for $\rho_{tr}$ based on the 
criterion of Woosley (2007) taking into account the 
effect of intermittency. Using the log-normal 
distribution to calculate the cumulative probability in eq. (7), 
we find that no regions of critical size that meet Woosley's 
criterion appear at density above $10^7$ g/cm$^3$. 
We cannot give an exact predicted transition density for this 
model because we do not have data at densities below 
$10^7$ g/cm$^{3}$ for relevant quantities listed in Table 1. 
Note that Woosley (2007) obtained a transition density 
around $10^7$ g/cm$^3$ under the assumption that a region of size $10^6$ cm with rms velocity of $10^8$ cm/s
is available. From our estimate above and the result in R\"opke (2007), the probability that such a region 
exists is small, therefore it is appropriate to take the transition density predicted in Woosley (2007) 
as an upper limit for his DDT criterion.        
 

\subsection{The log-Poisson model}
We next consider the log-Poisson model. 
Using the distribution eq (21), we have, 
\begin{equation}
\begin{array}{lll}
P (\epsilon_{l_c}>K^3\epsilon_f)& =& exp(-\lambda_c)    \sum \limits_{n=0}^{\infty}
\frac{\lambda_c^n}{n!} \int \limits_{ln(K^3\epsilon_f/\bar\epsilon)}^{\infty} P_L
(x-\gamma ln(L/l_c)- nln(\beta))dx\\
&=&exp(-\lambda_c)    \sum \limits_{n=0}^{\infty}
\frac{\lambda_c^n}{n!}\int \limits_{ln(K^3\epsilon_f/\bar\epsilon)-\gamma ln(L/l_c)- nln(\beta)} ^{\infty} P_L(x)dx\\
&=&exp(-\lambda_c)    \sum \limits_{n=0}^{\infty}
\frac{\lambda_c^n}{n!} F_n
\end{array}
\end{equation} 
where $\lambda_c= \gamma ln (L/l_c)/(1-\beta)$ and the integrals in the second 
line are denoted as $F_n$ for convenience.    
Note that the integral lower limit 
$ln(K^3\epsilon_f/\bar\epsilon)-\gamma ln(L/l_c)- nln(\beta)$ 
increases with $n$ because $\beta<1$, therefore $F_n$ is a decreasing function 
of $n$. An exact calculation for the cumulative probability is impossible because of the 
unspecified function $P_L$.  
We will neglect all the $n \ge 1$ terms and only keep the $n=0$ term in 
our calculation, i.e., we only include the contribution of the most 
intensive structures at scale $l_c$. Obviously, this approximation gives a 
lower limit for the probability and the transition density we obtain will also be 
a lower limit. We will show that the criterion for DDT obtained from 
this approximation is exact if $P_L$ is a delta function.

The contribution from $n=0$ is 
$exp(-\lambda_c)=(\frac {l_c}{L})^{\gamma/(1-\beta)} F_0$. 
For $\gamma=2/3$ and $\beta=2/3$, it is equal to $(l_c/L)^2 F_0$ 
and $N_{l_c} \times P(\epsilon_{l_c}>K^3\epsilon_f) 
\geq 4 \pi (R_f/l_c)^2 (l_c/L)^2 F_0= 4 \pi (R_f/L)^2 F_0$. Since the size of the flame region  $R_f \geq L$, 
it means that the number of regions which are larger than the 
critical size and in the distributed regime is $\simeq 4 \pi F_0$.  
Since the distribution $P_L(x)$ is probably strongly concentrated at $x=0$,  
the sufficient and almost necessary condition for $F_0 \simeq 1$ is that 
the integral limit $ln(K^3\epsilon_f/\bar\epsilon)-\gamma ln(L/l_c) \le 0$, 
or equivalently,    
\begin{equation}
\bar \epsilon > (l_c/L)^{2/3} K^3\epsilon_f
\end{equation}
which is a convenient criterion for the DDT in the log-Poisson model. 
Note this criterion is much weaker than the mean criterion eq (2). 
Once the condition is satisfied, at least one region of critical size that 
covers the flame enters the distributed regime due to the most intense 
stretching strength available at scale $l_c$.  

As mentioned in \S 3.2, if the dissipation structures are 2-dimensional, 
$\beta=1/3$. In that case, the contribution from the $n=0$ term is 
$(l_c/L) F_0$ and $N_{l_c} \times P(\epsilon_{l_c}>K^3\epsilon_f) \geq 4 \pi (R_f/l_c) (R_f/L) F_0$, 
which is much larger than $1$ if $F_0 \gsim 1$. Therefore, 
the criterion eq (25) is a sufficient condition for the case with 
sheet-like dissipation structures such as in MHD turbulence or 
highly compressible turbulence.  
    
We have neglected the $n>1$ terms in eq. (24), the contribution of which 
depends on how rapidly $P_L(x)$ decreases with $x>0$. We consider the 
extreme example where $P_L$ is a delta function. In this case, 
before the condition eq (25) is met, $F_n=0$ for any $n$ thus the cumulative probability is zero.  
When the condition is just satisfied as the density decreases,  
only the $n=0$ term contributes and all the $n>1$ terms are still zero, 
i.e., the most intensive ($n=0$) structures at $l_c$ can stretch a 
local flame into the distributed regime while all the less intensive 
structures ($n\ge 1$) still cannot.         
From the calculation above, we see that in this case once 
the $n=0$ term contributes, at least one region around the flame front 
experiences the largest stretching rate and enters the distributed regime.
Therefore, if $P_L$ is a delta function, eq (25) is  
both the necessary and the sufficient condition.  
This is true for both $\beta=2/3$ and $\beta=1/3$. 
If $P_L$ is not a delta function, the tail of $P_L$ gives rise to the 
possibility that the distributed regime can emerge in a local region 
of critical size before the condition eq (25) is met. This could lead 
to an even weaker condition than eq (25). Since we expect 
that $\epsilon_L$ can only vary within a factor of a few, 
the condition can be weaker only by a factor of a few. Because 
the r.h.s of the condition (23), especially $\epsilon_f$, depends on 
the density very sensitively, this would not increase the 
predicted $\rho_{tr}$ considerably.


The condition eq (25) can be easily applied to calculate the transition 
density using $\epsilon_f$ and $l_c$ in Table 1.   
For example, we get $\rho_{tr}= 8.7 \times 10^7$ g/cm$^3$   
for $U=100$ km/s and $L=100$ km if $K=1$. 
This result is consistent with that 
from the log-normal model and is also about a factor of 2 larger than the 
prediction by the mean criterion eq (2).  

The transition density predicted by the log-Poisson model with different 
parameters for turbulence is given in the 4th line of Table 2. 
Again the numbers in parenthesis are for $K=8$. The results are consistent 
with those from the log-normal model  
and are at least $2-3$ times larger than from the criterion eq (2).

Again we consider the possibility that there exists a region of size $10^6$ cm with rms turbulent 
velocity of $10^8$ cm/s required for DDT by Lisewski et al. (2000). 
The probability $P(\epsilon_{\rm 10^6 \hspace {1mm} cm}>10^{18} \hspace{1.5mm} 
{\rm cm^2/s^3})$ depends on $P_L$, the probability distribution of the dissipation rate at the integral scale $L$. 
Since $P_L$ is probably not universal and is flow-dependent, the log-Poisson model cannot give 
an exact estimate for the probability. Here we assume $P_L$ is a delta function and see under what 
condition it is possible to find a required region. We find that
the necessary and sufficient condition to have such a region is that $U > 10^{22/3} (L/{\rm cm})^{1/9}$ cm/s. 
For $L \simeq  10^7$ cm, $U$ has to be larger than $10^8$ cm/s. This can be understood from the fact that, 
in the log-Poisson model, the available kinetic energy in the most intermittent structures for dissipation is 
assumed to be the kinetic energy at the integral scale (see Appendix A). Since $U>10^8$ cm/s is probably not 
achievable, it is rare that a region as required by Lisewski et al. (2000) exists, again in agreement with 
R\"opke (2007).    
          
Using the same calculation that leads to eq (25), we obtain a criterion 
for the DDT model by Woosley (2007) accounting for the intermittency effect, 
\begin{equation}
\bar \epsilon >(l_c/L)^{2/3}(l_c/l_f)^2 \epsilon_f.
\end{equation}
This is weaker than the corresponding mean criterion eq (6) by 
a factor of $(l_c/L)^{2/3}$, meaning that intermittency increases the transition density. In comparison with 
eq (25), the condition is stronger and thus gives a smaller transition density than that for the KOW and 
NW model with intermittency included.  
At smaller density, the critical length $l_c$ is larger and the factor $(l_c/L)^{2/3}$, representing the
intermittency effect, is closer to unity. This implies that intermittency gives a weaker effect on the
the transition density for the Woosley (2007) criterion than for that by KOW and NW. 
Using Table 1, we again find that the condition eq (26) is not satisfied at densities above 
$10^7$ g/cm$^3$ for the five cases listed in Table 2, i.e., the predicted transition density is still below 
$10^7$ g/cm$^3$ after including intermittency (see discussion in \S 4.1).   
\\
\\

In summary, intermittency can considerably enhance the onset of the 
distributed flame regime and hence increase the transition density in 
the DDT model of KOW and NW. Both the intermittency models we consider 
here predict a transition density $2-3$ times larger than from the criterion 
using the mean dissipation rate. This factor of $2-3$ brings the 
transition density to be in disagreement with the observational constraints 
for turbulent velocity larger than $U=10^6$ cm/s in the case $K=1$. 
We discuss the implications of this result in the next section.      
We also find that existence of regions of size $10^6$ cm with velocity $10^8$ cm/s is
rare,  in agreement with the numerical result of R\"opke (2007). The strong DDT criterion 
given by Woosley (2007) gives a transition density below $\rho_{tr}=10^7$ g/cm$^{3}$ even when 
intermittency is included. We expect that the intermittency effect is weaker for stronger 
DDT criteria.

\section{Conclusion and Discussion}
We have studied the effect of intermittency on the transition from the 
flamelet regime to the distributed regime in Type Ia SNe, and hence
on the transition density for the DDT model by KOW and NW.
In their model, the detonation occurs via the Zel'dovich mechanism that
requires a nearly isothermal region larger than a critical size to drive a 
sufficiently strong supersonic shock. KOW and NW assumed that 
the almost isothermal mixture of fuel and ash can be produced 
once turbulence is strong enough to get the flame into the distributed regime. 
The DDT is assumed by KOW and NW to occur when the {\it average} flow gets 
into the distributed regime. We argue that  
the sufficient condition for the DDT is that there is {\it one} region that is 
larger than the critical size and in the distributed regime.   

The intermittency in turbulence, as a result of the spatial inhomogeneity 
of the dissipation rate, gives rise to regions with strong local turbulent 
strength that can force the flame into the distributed regime earlier than 
elsewhere. Therefore the transition from the flamelet regime to the 
distributed regime is not spatially smooth, but intermittent. 
At early time when the density in the white dwarf is large, the flame has 
a large speed and a small width and thus resists being 
efficiently stretched and broken by the turbulence.  
At the same time, the critical size is very small. This has two effects 
that tend to make an early DDT likely. First, the intermittency of 
turbulence tells us that the probability of finding extremely strong 
stretching within a smaller critical size is larger. Second, there are more 
regions of smaller sizes available.  
Therefore it is possible that the DDT is triggered at a small ``spot'' 
when the density is larger than needed for the average flow to enter the 
distributed regime. As we pointed out in the Introduction, the critical size 
as a function of the density plays an important role in determining the 
transition density for the DDT in our calculations.       


We used two analytical intermittency models to statistically investigate
when the first region appears which is both larger than the critical size 
and in the distributed region. This is assumed to be the time when the DDT 
occurs by KOW and NW. We found that, for various parameters for the intensity and length 
scales, DDT occurs at a transition density at least $2-3$ times larger than 
the density at which the average flow enters the distributed regime. 
The transition density has been determined empirically by invoking it as a 
free parameter in spherically-symmetric models and then computing models 
that best match the observed multicolor light curve shapes and magnitudes
(H\"oflich \& Khokhlov 1996). 
Recognizing that the spherical models are oversimplified,
they do give some guidance to the empirical constraints
on the density at which DDT occurs.
H\"oflich (1995) used this procedure to fit 
observations of the Branch core normal SN 1994D and preferred a value of the 
transition density of $2\times10^7$ gm cm$^{-3}$. H\"oflich, Khokhlov \&
Wheeler (1995) explored a range of transition densities in the context of 
pulsating delayed detonation models and favored densities in the range 
$0.8 - 2.2\times10^7$ gm cm$^{-3}$. Dominguez, H\"oflich \& Straniero (2001)
adopted $2.3\times10^7$ gm cm$^{-3}$. Allowing for an uncertainty of 
a factor of 2, the predicted transition densities by the mean criterion 
are consistent with $2 \times10^7$ gm cm$^{-3}$ as favored by the 
observations in all the cases except that 
with $\bar \epsilon = 10^{16}$ cm$^2$/s$^3$ and $K=1$ (Table 2). 
With the intermittency effects we have examined here, the transition density 
would be a factor of $2-3$ higher. If $K=1$, all the predicted $\rho_{tr}$ are 
larger than $2\times10^7$ gm cm$^{-3}$ by at least a factor of $2$ 
except the case with $U=10^6$ cm/s. The predicted transition density with 
$K=8$ is $2-3$ times smaller than from $K=1$. From Table 2, the predicted 
$\rho_{tr}$ for the intermittency models with $K=8$ agree with the 
observations within a factor of $2$ except the case with a large velocity 
scale $U=10^8$ cm/s at the integral length scale. 
To avoid discrepancy with the observations, 
our result indicates several possibilities.  
\begin{enumerate}  
\item 
The large scale motions caused by Rayleigh-Taylor instability 
freeze out due to the overall expansion of the star (Khokhlov 1995). The 
freezeout effect has to be efficient enough so that the developed part of
the flow has a velocity scale of $\lsim 10^6$ cm/s (see Table 2). 
\item 
The flame is very robust. To break the flame, the local Gibson scale has to 
be at least $K^3=512$ times smaller than the flame thickness. In this case, 
the predicted transition density is $2-3$ times smaller than from $K=1$. 
\item     
There is not enough time for the buoyancy-driven turbulence 
to fully develop down to the critical size before the predicted density 
for the DDT by our intermittency models is attained, thus motions at scales 
below the critical size are either absent or non-intermittent.   

\item
The DDT does not occur immediately after a region of the critical size 
enters the distributed regime. It may take some time for turbulence to help 
mix the region and make it nearly isothermal. However, the time scale  
for turbulence to mix a region of the critical size in the distributed 
regime is very small, $\lsim 10^{-2}$ s at densities larger than 
$3 \times 10^7$ g/cm$^3$. It is unlikely that the density 
drops much in such a short timescale.  

\item 
Having a large enough region entering the distributed regime is 
not a sufficient condition for detonation. 
As mentioned in the Introduction, there are several uncertainties in the simple 
model by KOW and NW assuming flame 
quenching, entering the distributed regime and the DDT all occur simultaneously.  
\end{enumerate}

Our result alludes to the possibility that the criterion by KOW and NW is too weak
for the DDT, supporting the claim of Lisewski et al (2000) and Woosley (2007) that
just entering the distributed regime is not sufficient for the DDT. We have shown that 
their criteria for the DDT are much stronger than just entering the distributed regime. 
We also studied the intermittency effect on their conditions for the DDT. We find 
that the existence of a region of size 1$0^6$ cm with rms turbulent velocity of $10^8$ cm/s 
required by Lisewski et al (2007) for a DDT is rare, consistent with numerical results of 
R\"opke (2007). We have also examined the intermittency effect on the transition 
density for the DDT criterion by Woosley (2007). We find that the effect is weaker for 
the stronger criterion and does not increase $\rho_{tr}$ to above $10^7$ g/cm$^{3}$. 
Woosley (2007) obtained $\rho_{tr}$ around $10^7$ g/cm$^{3}$ because of the assumption of 
strong turbulence velocity of $10^8$ cm/s in a region of size $10^6$ cm. Since the 
existence of such a region is rare, Woosley (2007) may considerably overestimate the transition 
density. This may imply that the condition for DDT by Lisewski et al. (2000) and Woosley 
(2007) is too strong and predicts a transition density smaller than that empirically determined  
from observations.             

We have only studied the DDT in white dwarfs with the 
initial chemical composition of half carbon and half oxygen. 
In a white dwarf with more carbon, the nuclear timescale is smaller 
thus a stronger turbulent intensity is needed to break the flame. 
This results in a smaller transition density using the mean criterion (NW). 
On the other hand, the critical size for detonation in 
such a white dwarf is smaller, therefore the intermittency effect could be 
more efficient in increasing the transition density. We were not 
able to perform a calculation for the chemical composition with the 
carbon abundance larger than 0.5 due to the lack of sufficient 
data for the critical length scale in this case. 


We point out that the intermittency models we used were originally proposed 
for homogeneous and isotropic turbulence. Turbulence in SNe Ia is stratified 
and may not achieve homogeneity and isotropy at very small scales even 
if the turbulence is developed at these scales. The effect of the departure 
from homogeneity and isotropy on the predicted transition density is out of 
the scope of this paper.

\acknowledgements
We thank Elaine Oran for useful discussions. This research was supported in 
part by NSF Grant AST-0707769 (LP, JCW) and by NASA Astrophysics Theory 
Program Grant NAG5-13280 (JS).
 

\appendix
\section{The She-Leveque Model}
In their original paper, She and Leveque (1994) start with the moments of the distribution $P(\epsilon_l)$ and use the ratios of two successive moments, $\epsilon_l^{(p)} = \langle \epsilon_l^{p+1} 
\rangle/\langle \epsilon_l^{p} \rangle \propto l^{\tau_{p+1}-\tau_p}$, to characterize a hierarchy of 
dissipative structures. This ratio can be written as 
$\epsilon_l^{(p)} =\int \epsilon_l Q_p(\epsilon_l) d\epsilon_l $ where $Q_p(\epsilon_l)=\epsilon_l^{p} P(\epsilon_l)/\int \epsilon_l^{p} P(\epsilon_l) d\epsilon_l$. For a typical distribution $P(\epsilon_l)$ that decreases monotonically and faster than any power law at large $\epsilon_l$, 
$Q_p(\epsilon_l)$ strongly peaks around $\epsilon_l^{(p)}$ for large $p$. 
Clearly $\epsilon_l^{(p)}$ increases with $p$, and $\epsilon_l^{(\infty)}=\lim_{n\to\infty} \langle \epsilon_l^{p+1} \rangle/\langle \epsilon_l^{p} \rangle$ corresponds to the most intense dissipative structures at scale $l$. These strongest dissipative structures are the origin of the anomalous scaling and 
 the scaling of $\epsilon_l^{(\infty)}$ with $l$,
 \begin{equation}
\epsilon_l^{(\infty)} \sim l^{-\gamma} 
\end{equation}  
is of fundamental importance. To determine the parameter $\gamma$, we can dimensionally write $\epsilon_l^{(\infty)}$ as
an energy scale divided by a time scale $t_l$. She and Leveque argued that for the most intermittent structures this energy scale is the largest 
available kinetic energy (which is $\sim v_{rms}^2$, independent of $l$) and assumed that $t_l$ exhibits 
a regular Kolmogorov scaling $t_l \sim l^{2/3}$, therefore $\gamma=2/3$.  
From eq (A1), we have  $\tau_{p+1}=\tau_p-\gamma$ for $p \to \infty$ or
\begin{equation}
\tau_p=-\gamma p +C, \hspace{3mm} p \to \infty
\end{equation}
where the constant $C$ has a physical interpretation as the codimension of the most intermittent structures.  Eq (A1) means that the dissipation rate in a region of size $l$ that encloses the most intensive structures scales with $l$ as $\epsilon_l \propto l^{-\gamma}$.  
When calculating $\langle \epsilon_l^p \rangle$ at $p \to \infty$, we need to consider  the possibility of a point finding itself within a distance $l$ to the most intermittent structures, which is 
proportional to $l^{(D-d)}$ where $D$ is the dimension of the system and 
$d$ is the dimension of the most intensive structures (Frisch 1995). 
As $p \to \infty$, the contribution to $\langle \epsilon_l^p \rangle$ is dominated by the most intermittent structures,  therefore $\langle \epsilon_l^p \rangle \propto l^{-\gamma p +D-d}$. 
Comparing with eq (A2), we find that $C$ corresponds to the codimension of the most intermittent structures, $C=D-d$. 

In order to determine the entire hierarchy of the dissipative structures, 
She and Leveque argued  that the intensity $\epsilon_l^{(p+1)}$ of the dissipative structures at level 
$p+1$ depends only on their immediate precursor,  the structures of level $p$, from which the level $p+1$ structures directly develop, and on the most intensive structures, where the structures of all orders 
tend to end up.  Based on this argument, they made an assumption about the hierarchy of the dissipation rates,   
\begin{equation}
\epsilon_l^{(p+1)}=A_p (\epsilon_l^{(p)})^{\beta}(\epsilon_l^{(\infty)})^{1-\beta}
\end{equation}   
where the coefficients $A_p$ are independent of $l$ but may be flow-dependent and non-universal.  
The parameter $\beta$ will be completely fixed by $\gamma$ and the codimension $C$. According to
She and Leveque, this relation corresponds to a mysterious symmetry of the Navier-Stokes equation, termed  "the hidden symmetry".   

To derive $\tau_p$ from equation (A3), it is convenient to define a new variable,  
\begin{equation}
\pi_l=\epsilon_l/\epsilon_l^{(\infty)}  
\end{equation}
which was introduced by Dubrulle (1994). Clearly  $\epsilon_l^{(p)}/\epsilon_l^{(\infty)}=
\langle \pi_l ^{p+1}\rangle/ \langle \pi_l ^{p}\rangle$ from the definition of $\epsilon_l^{(p)}$. 
Then the ``hidden symmetry'' assumed by She and Leveque becomes, 
\begin{equation}
\frac {\langle \pi_l ^{p+2}\rangle}{ \langle \pi_l ^{p+1} 
\rangle}= A_p (\frac {\langle \pi_l ^{p+1} 
\rangle}{\langle \pi_l ^p \rangle})^\beta  
\end{equation}
This recursion relation is solved by, 
\begin{equation}
\frac{\langle \pi_l ^{p+1} \rangle}{ \langle \pi_l ^p \rangle}=
C_p \langle \pi_l \rangle^{{\beta}^p} 
\end{equation}
where $C_0=1$ and $C_p=\prod \limits_{n=0}^p A_n^{\beta n}$ for $p>0$.
Eq. (A6) gives,  
\begin{equation}
\langle \pi_l ^{p} \rangle = B_p {\langle \pi_l \rangle} 
^{(1-\beta^p)/{1-\beta}}    
\end{equation}
where $B_0=B_1=1$  and $B_p=\prod \limits_{n=1}^{p-1} C_n$ for $p>1$. 
Noting that $\langle \pi_l \rangle=\bar{\epsilon}/ \epsilon_l^{(\infty)} \propto (\epsilon_l^{(\infty)})^{-1}$ and 
$\langle \pi_l^p \rangle = \langle \epsilon_l^p \rangle (\epsilon_l^{(\infty)})^{-p}$, eq (A7) gives
$\langle \epsilon_l^p \rangle \propto (\epsilon_l^{(\infty)})^{p- (1-\beta^p)/(1-\beta)}$. 
Using eq (A1), we have 
\begin{equation}
\tau_p=-\gamma p+ \gamma \frac{(1-\beta^p)}{1-\beta}
\end{equation}
The parameter $\beta$ is determined by the asymptotic behavior of $\tau_p$ at the $p \to \infty$, eq (A2). 
Letting $p \to \infty$ in eq (A8) and comparing with eq (A2), we find 
that $\gamma/(1-\beta)=C=D-d$. Since the most intermittent structures in 3D incompressible turbulence are filamentary, we have $d=1$ and $C=2$, thus $\beta=1-\gamma/2=2/3$ for $\gamma=2/3$.  
Finally we arrive at the celebrated She-Leveque formulae, 
\begin{equation} 
\tau_p=-2p/3+  2(1-(2/3)^p) 
\end{equation}
and 
\begin{equation} 
\zeta_p=p/9+  2(1-(2/3)^{p/3}) 
\end{equation}
which agrees with the experimental result with an accuracy of $1\% $.
Note that this result is consistent with the Kolmogorov's exact result for the third order structure function, i.e., $\zeta_3=1$. 
If the most intense structures are two dimensional, e.g., the dissipation sheets in compressible flows 
(Boldyrev et al. 2002) or the current sheets in MHD (Muller \& Biskamp 2000), $\beta=1/3$.

\clearpage

\begin{table}
\begin{center}
\caption{The laminar flame speed, the flame thickness and the critical length for a white 
dwarf with half carbon and half oxygen
\label{tbl-1}}
\begin{tabular}{ccccc}
\tableline\tableline
$\rho$ ($10^9$ g/cm$^3$) & $S_l$ ($10^5$ cm/s) & \ $l_f$ (cm) &  
$\epsilon_f $ ($10^{15}$ cm$^2$/s$^3$)&$l_c$(cm)\\
\tableline
2 & 75.8 & 9.35(-5) & 4.66(9) & 7(1)\\
0.5 & 18.1 & 9.46(-4) &  6.27(6) & ---\\
0.1 & 2.33 & 2.75 (-2) & 4.60(2) & 2(2)\\
0.05 & 0.599 & 5.19 (-1) & 0.414 & 1.3(3){\tablenotemark{a}}\\
0.03 & 0.26{\tablenotemark{b}}  & 1.78{\tablenotemark{b}} & 0.98 (-2) &5(3)\\
0.01 & 4.72(-2)& 4.22& 2.59(-5) & 2(5)\\ 
\tableline
\end{tabular}
\tablenotetext{a}{Read from Fig. 6 in KOW.}
\tablenotetext{b}{Read from Fig. 7 in KOW.}
\tablecomments{The values of $S_l$ and $l_f$ are mainly taken from Table 3 of 
Timmes and Woosley 1992. The value of $l_c$ is 
mainly taken from NW. We also include numbers (marked) from KOW because their 
results are very similar to $NW$ despite the difference of details 
in the two models.   
Numbers in parentheses are powers of 10. } 
\end{center}
\end{table}

\begin{table}
\begin{center}
\caption{The predicted transition densities $\rho_{tr}$ (in unit of 
$10^7$ g/cm$^3$) for various models
\label{tbl-2}}
\begin{tabular}{cccccc}
\tableline\tableline
Cases {\tablenotemark{a}} & A & B & C & D & E\\
\tableline
mean criterion & 6.9(3.3) & 5.5(2.3) & 4.1(1.5) & 3.0(1.0) 
& 2.0 (---)\\
log- normal {\tablenotemark{b}}& 27(10) & 11(4.9) & 10(4.4) & 9.4 (4.0) 
& 4.3 (---)\\
log- normal {\tablenotemark{c}}& 23(7.6) & 9.7(4.9) & 8.4(3.9) & 7.1 (3.0) 
& 4.3 (---)\\
log- Poisson & 24(8.6) & 9.5(4.3) & 8.7(3.8) & 7.9(3.3) 
& 3.8 (---)\\
\tableline
\end{tabular}
\tablenotetext{a}{A: $U=10^8$ cm/s, $L=10^8$ cm, $\bar\epsilon
=10^{16}$cm$^2$/s$^3$;\\
B: $U=10^7$ cm/s, $L=10^6$ cm, $\bar\epsilon=10^{15}$cm$^2$/s$^3$;\\
C: $U=10^7$ cm/s, $L=10^7$ cm, $\bar\epsilon=10^{14}$cm$^2$/s$^3$;\\
D: $U=10^7$ cm/s, $L=10^8$ cm, $\bar\epsilon=10^{13}$cm$^2$/s$^3$;\\ 
E: $U=10^6$ cm/s, $L=10^6$ cm, $\bar\epsilon=10^{12}$cm$^2$/s$^3$ }
\tablenotetext{b}{The predicted transition density 
assuming a perfect fit of the distribution $P(\epsilon_{l_c})$  
by log-normal. This is the upper limit for $\rho_{tr}$ since the 
log-normal approximation may break down and overestimate the 
distribution at the far tail.}
\tablenotetext{c}{The lower limit for the transition density assuming 
the log-normal approximation applies only up to $5-\sigma$.}
\tablecomments{The numbers in parentheses are the results predicted if the 
Gibson scale has to be 512 times smaller than the flame thickness for the 
transition to the distributed regime. In case E, (---) indicates 
that the transition density is smaller than $10^7$ g/cm$^3$, 
which cannot be well estimated since we only have data 
to $10^7$ g/cm$^3$ in Table 1.} 
\end{center}
\end{table}

\end{document}